\documentclass{kluwer}    
\usepackage[]{graphicx}
\newdisplay{guess}{Conjecture}

\begin{document}                                                                                   
\begin{article}
\begin{opening}         
\title{On the Origin of High-eccentricity Halo Stars} 
\author{C.B. \surname{Brook}}
\author{D. \surname{Kawata}} 
\author{B.K. \surname{Gibson}}
\author{C. \surname{Flynn}}
\institute{Centre for Astrophysics \& Supercomputing, 
Swinburne University, Australia}                               
\runningtitle{On the Origin of High-eccentricity Halo Stars}
\runningauthor{Brook, Kawata, Gibson, Flynn}

\begin{abstract}
The present-day chemical and dynamical properties of the Milky Way are
signatures of the Galaxy's formation and evolution.  Using a self
consistent chemodynamical evolution code we examine these 
properties within the currently favoured paradigm for galaxy formation - 
hierarchical clustering within a CDM cosmology.
Our Tree N-body/Smoothed
Particle Hydrodynamics code includes a self-consistent treatment of
gravity, hydrodynamics, radiative cooling,
star formation, supernova feedback and chemical enrichment.  
Two models are described which
explore the role of small-scale density perturbations in driving
the evolution of structure within the Milky Way. The relationship between
metallicity and kinematics of halo stars are quantified and the implications
for galaxy formation discussed.  While high-eccentricity
halo stars have previously been considered a signature of 
``rapid collapse'', we suggest
that many such stars may have come from recently accreted satellites.
\end{abstract}
\keywords{Milky Way, galaxy formation, Galactic halo}

\end{opening}           

\vspace{-0.4cm}
\section{Introduction} 
\vspace{-0.4cm}
The ``monolithic collapse'' versus ``satellite accretion'' debate
surrounding galaxy formation is a classic one, and one which received 
attention once again at this Euroconference III.
The former scenario was best expressed by Eggen,
Lynden-Bell \& Sandage (1962, hereafter ELS); supporting evidence for
the ELS picture came from the apparent
positive correlation between eccentricity and
metallicity of halo stars. However, current cosmological theories of structure
formation have more in common with accretion-style
scenarios like that envisioned by Searle \& Zinn (1978). 
Evidence in support of the latter can be found in the observations of
stellar phase space substructure in the Galactic halo 
(e.g. Helmi et~al. 1999).

We were motivated to run a grid of chemodynamical simulations with the
intention of contrasting the effects of the two collapse scenarios on the
evolution of the Milky Way.  The two models described here
vary primarily in their degree of
clustering, and we examine the properties of the resulting
simulated galaxies, in order to uncover present-day ``signatures'' of
the model initial conditions and evolution.  Here, we focus on the
distribution of halo star orbital eccentricities.

\vspace{-0.4cm}
\section{The Code and Models}
\vspace{-0.4cm}
Our Galactic ChemoDynamical code ({\tt GCD+})
models self-consistently the effects of
gravity, gas dynamics, radiative cooling, and star formation. Type~Ia and
Type~II supernova feedback is included.  We relax the instantaneous recycling
approximation when monitoring the Galactic chemical evolution.
Details of {\tt GCD+} can be found in Kawata \& Gibson (2003, in prep);
an earlier version of the code is described in Kawata (2001).
  
The semi-cosmological version of {\tt GCD+} used here is based upon the code of
Katz \& Gunn (1991).  The initial
condition is an isolated sphere of dark matter and gas, onto which  small scale
density fluctuations are superimposed (parameterised by $\sigma_8$).  These
perturbations are the seeds for local collapse and subsequent star formation.
Solid-body rotation is imparted to the initial sphere; 
this determines whether
a disk-like or elliptical galaxy results.  For the models described here,
relevant parameters include
the total mass ($5\times 10^{11}$~M$_\odot$), baryon fraction
($\Omega_{b}=0.1$), and spin parameter ($\lambda=0.0675$); we employed
38911 dark matter and 38911 gas/star particles.

Again, the two models described here differ only in the value of
$\sigma_8$. In model~1,
$\sigma_8=0.5$, as favoured in standard CDM
($\Omega_0 = 1$) cosmology. In model~2, $\sigma_8
=0.04$, a smaller value which results in a more rapid, dissipative, collapse.

\begin{figure} 
\hspace{-2pc}
\includegraphics [width=30pc]{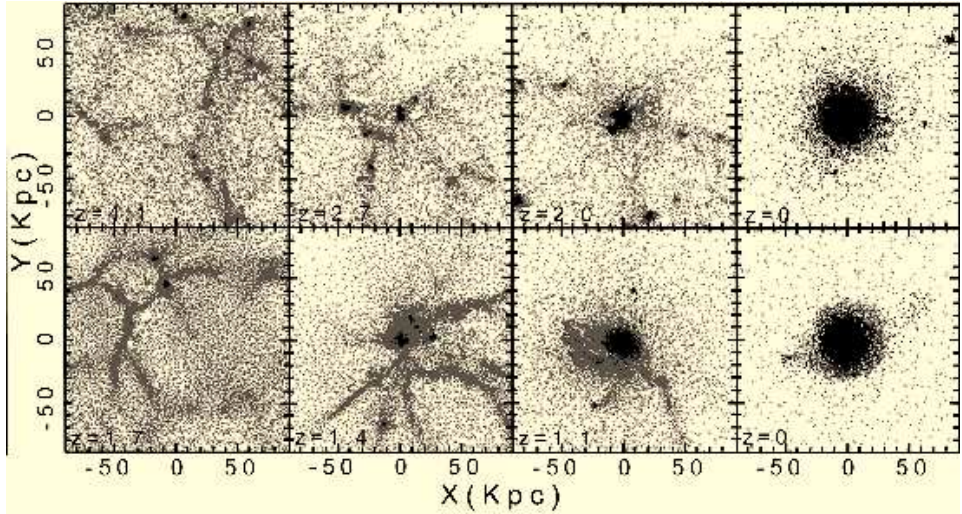}
\caption[]{$x-y$ plots of model 1 (upper panels) and model 2 (lower panels).
The $z$ axis is the initial rotation axis. Grey dots represent
gas particles, while black represent star particles.
Epochs are chosen so that roughly the same stellar
mass is present in corresponding upper \& lower panels. Gas collapse and star
formation are more centralised in model 2.}
\label{chibadat}
\end{figure}

\section{Results}
\vspace{-0.4cm}
Figure~1 demonstrates the classical hierarchical merging in action in both 
models 1 (upper panels) and 2 (lower panels).  Gas particles are marked in grey, while
star particles are in black.  Star formation occurs in
overdense regions, seeded by the initial small-scale perturbation
spectrum.  Stars continue to form in sub-clumps, as well
as in the central region of the disk galaxy as it is built up.  We see
less clustering in model 2 with most of the star formation occurring in the
central region of the galaxy.

We analysed the bulk properties of the models at $z$=0 and confirmed that
they were consistent with those
of Berczik (1999) and Bekki \& Chiba (2001).
The predicted surface density
profiles, metallicity gradients, and rotation curves for our two models did
not differ significantly.  However, we did find a difference in the
distribution of the eccentricities of the orbits of solar neighbourhood
halo stars.

\begin{figure}
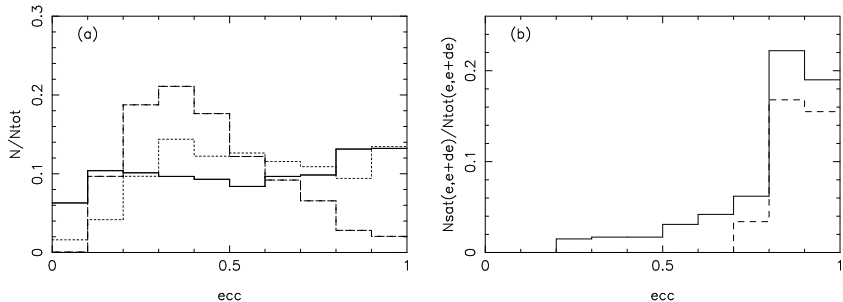

\begin{tabular}{c@{\hspace{1pc}}c}
\includegraphics[width=2.1in]{brook.fig2a.ps} &
\includegraphics[width=2.1in]{brook.fig2b.ps}
\end{tabular}
\caption{$(a)$ The eccentricity distribution of halo stars at the solar
circle for our two models. The dotted (dashed) line shows model 1 (model
2). The solid line shows observations of Chiba \& Beers (2000). Model 1
produced a greater number of high eccentricity
halo stars in the solar neighbourhood. \hspace{8cm} $\qquad (b)$ The solid
line shows the eccentricity distribution of solar neighbourhood halo stars
which were in satellites at $z$=0.46. The dashed line shows stars
originating from a single satellite. The $y$-axis is normalised by the total
number of stars in each eccentricity bin from Figure 2a.}
\end{figure}

The histogram of Figure~2a shows the eccentricity distribution of halo
star particles ([Fe/H]$<$$-$0.6) in the solar neighbourhood for the two
models. Each bin shows the fraction of such star particles falling in a
given eccentricity range. Also shown are observations from Chiba \& Beers
(2000, hereafter CB).  Model 1 produced
a greater number of high eccentricity (ecc $>$ 0.8) solar neighbourhood
halo stars, and is in better agreement with observation.

We next examined the specific accretion history of each model, tracing the
eccentricity distribution functions for the stars associated with each
disrupted satellite.  
We identified satellites at $z$=0.46 which have merged 
into the halo of the host galaxy by $z$=0.  
The histogram of Figure 2b shows the eccentricity distribution 
of solar neighbourhood halo
stars which originated in these recently accreted satellites. The $y$-axis is
normalised by the total number of solar neighbourhood halo stars in each
eccentricity bin. 
Our primary conclusions are that the majority of these halo stars are of
high-eccentricity, and that one satellite in particular contributes
$\sim$20\% of
all high eccentricity halo stars in the solar neighbourhood at $z$=0.
The reader is directed to the complementary study of Steinmetz et~al. (these
proceedings) which finds that stars from accreted satellites which were on polar orbits form part of the galaxies halo.  

\vspace{-0.4cm}
\section{Conclusions}
\vspace{-0.4cm}
The key question we wish to address remains ...
\it what are the implications for the competing galaxy formation paradigms? \rm

A brief response is as follows: CB observationally found no correlation between eccentricity and metallicity for halo stars near the Sun (their Figure 6a), obviating the need for a ``rapid collapse'' picture of the formation of the Galaxy (ELS). However, CB do identify a clump of 
high-eccentricity low-metallicity ([Fe/H]$\sim$$-$1.7) stars in this
observational plane.  In terms of ELS, they
interpret this clump as a relic of a rapid collapse phase. Our
simulations suggest that this clump is, more likely, evidence of recent
satellite accretion in the Galactic halo.

\vspace{-0.4cm}
\acknowledgements
\vspace{-0.4cm}
CBB thanks the Organising Committee for financial assistance, and
appreciates the hospitality of Gerhard Hensler and Andi Burkert in
arranging collaborative visits to Kiel and MPIfA, Heidelberg.
BKG acknowledges the financial support of the Australian Research
Council through its Large Research Grant Program (\#A00105171).
We acknowledge
the generous support of the Australian Partnership for Advanced Computing
through its Merit Allocation Scheme.

\end{article}

\vspace{-0.4cm}
\begin{thebibliography}{}
\vspace{-0.4cm}
\bibitem[\protect\citeauthoryear{BC}{2001}]{BC}
Bekki, K. \& Chiba, M. 2001,
ApJ, 558, 666

\bibitem[\protect\citeauthoryear{Berczik}{1999}]{}
Berczik, P. 1999,
A\&A, 348, 371

\bibitem[\protect\citeauthoryear{CB}{2000}]{CB}
Chiba, M. \& Beers, T.~C. 2000,
ApJ, 119, 2843  (CB)

\bibitem[\protect\citeauthoryear{ELS}{1962}]{ELS}
Eggen, O.~J., Lynden-Bell, D. \& Sandage, A.~R. 1962,
ApJ, 136, 748  (ELS)

\bibitem[\protect\citeauthoryear{Helmi}{1999}]{Helmi}
Helmi, A., White, S.~D.~M., deZeeuw, P.~T. \& Zhao, H.S. 1999,
Nature, 402, 53 

\bibitem[\protect\citeauthoryear{Kawata}{2001}]{Kawata}
Kawata, D. 2001,
ApJ, 558, 598 

\bibitem[\protect\citeauthoryear{KG}{1991}]{KG}
Katz, N. \& Gunn, J.E. 1991,
ApJ, 377, 365 

\end{thebibliography}
\end{document}